**Comments on Van der Waals supercritical fluid: Exact formulas for special lines [J. Chem. Phys. 135, 084503 (2011)]**

I.H. Umirzakov [a)]

*Institute of Thermophysics, Siberian Branch of Russian Academy of Sciences, Novosibirsk, 630090, Russia*

**Abstract**: The equations for Widom lines for the supercritical van der Waals fluid were considered in the paper J. Chem. Phys. 135, 084503 (2011). But the equations for the speed of sound and its special Widom line were incorrect. The correct equations are obtained.

The equations for "special lines" or Widom lines for the supercritical van der Waals fluid were considered in the paper.[1] Two errors of the paper were corrected by authors.[2] But there are another two drawbacks in the paper.[1]

According to the paper[1] the speed of sound is defined by the incorrect relation $V_s = \sqrt{T/\zeta_T}$, where $T$ is the temperature, $\zeta_T = T(\partial \rho / \partial P)_T$, $P$ is the pressure and $\rho$ is the density. The correct relation for the speed of the sound is[3]

$$V_s = \sqrt{C_P (\partial P/\partial \rho)_T / C_V} , \qquad (1)$$

where $C_P$ and $C_V$ are isobaric and isochoric heat capacities, respectively. Using exact thermodynamic relation $C_P = C_V + T(\partial P/\partial T)^2_\rho / \rho^2 (\partial P/\partial \rho)_T$ and the Van der Waals equation of state we have from Eq. (1)

$$V_s = \sqrt{kT(1 + k/C_{V,ig})/m(1 - b\rho/m)^2 - 2a\rho/m} , \qquad (2)$$

where $k$ is the Boltzmann's constant, $m$ is the mass of one molecule (atom), $C_{V,ig}$ is the isochoric heat capacity of the ideal gas per molecule (atom), which can depend only on the temperature, $a$ and $b$ are the parameters of the Van der Waals equation of state.

For the line of the minimum of the speed of sound on the isotherms we obtain from Eq. (2)

$$\rho_r = 3 - 2T_r^{1/3} \cdot (1 + k/C_{V,ig})^{1/3} ,$$

where $\rho_r = \rho/\rho_c$, $T_r = T/T_c$, $\rho_c = m/3b$ and $T_c = 8a/27b$. Therefore the statement of the paper[1] that the line of the minimum of the speed of sound on the isotherms obviously corresponds to the equation $\rho_r = 3 - 2T_r^{1/3}$, is incorrect.

[a] Electronic mail: umirzakov@itp.nsc.ru